\documentclass[aps,prl,superscriptaddress,showkeys,showpacs,twocolumn,nofootinbib,notitlepage,10pt]{revtex4-2}
\usepackage{amsmath,amssymb,slashed}
\usepackage[utf8]{inputenc}
\usepackage{graphicx}
\usepackage{hyperref}
\usepackage{bm,bbold}
\usepackage{accents}
\usepackage{ulem}
\usepackage{slashed}
\def\!{\mskip-\thinmuskip}
\newcommand{\di}{{\rm d}}

\newcommand{\wT}{\widehat{T}}

\newcommand{\wrho}{\widehat{\rho}}

\newcommand{\h}[1]{\widehat{#1}}
\newcommand{\tr}{{\rm tr}}  
\newcommand{\e}{{\rm e}}
\newcommand{\I}{{\rm i}}

\newcommand{\de}{\partial}

\renewcommand{\tilde}[1]{{\widetilde{#1}}}

\newcommand{\spt}{\cal{S}}
\def\wspt{{\widehat{\cal S}}}
\def\wPhi{{\widehat{\Phi}}}
\newcommand{\SP}{\mathfrak{S}}
\newcommand{\SPd}{\mathfrak{s}}

\usepackage[usenames,dvipsnames,svgnames,table]{xcolor}

\begin{document}
	%**************************************************************************************************
	\title{Emergent Canonical Spin Tensor in the Chiral-Symmetric Hot QCD}
	
	\author{Matteo Buzzegoli}\email{matteo.buzzegoli@e-uvt.ro}\affiliation{Department of Physics and Astronomy, Iowa State University, 2323 Osborn Drive, Ames, Iowa 50011, USA}
	\affiliation{Department of Physics, West University of Timișoara, Bd. Vasile Pârvan 4, Timișoara 300223, Romania}
	
	\author{Andrea Palermo}
	\email[]{andrea.palermo@stonybrook.edu}
	\affiliation{Center for Nuclear Theory, Department of Physics and Astronomy, Stony Brook University, Stony Brook, New York 11794-3800, USA}
	
	%**************************************************************************************************
	\begin{abstract}
		The spin tensor is fundamental to relativistic spin hydrodynamics, but its definition is ambiguous due to the pseudogauge symmetry. We show that this ambiguity can be solved in interacting field theories. We prove that the mean-field limit of a modified Nambu-Jona-Lasinio model with spin-spin interactions is equivalent to nondissipative spin hydrodynamics with a canonical spin tensor.
	\end{abstract}
	\maketitle
	
	%*********************************************************************************************************
	{\it{Introduction -}}
	%*********************************************************************************************************
	There is an ongoing intense debate concerning the proper definition of the spin tensor (ST), which is of fundamental importance in spin hydrodynamics~\cite{Becattini:2011ev,Becattini:2018duy,Becattini:2020riu,Buzzegoli:2021wlg,Speranza:2020ilk,Fukushima:2020ucl,Das:2021aar,Dey:2023hft,Florkowski:2018fap,Li:2020eon,She:2021lhe,Biswas:2023qsw,Florkowski:2017ruc,Florkowski:2017dyn,Florkowski:2018myy,Weickgenannt:2019dks,Bhadury:2020puc,Weickgenannt:2020aaf,Shi:2020htn,Bhadury:2020cop,Singh:2020rht,Peng:2021ago,Sheng:2021kfc,Weickgenannt:2022qvh,Gallegos:2020otk,Garbiso:2020puw,Montenegro:2017rbu,Montenegro:2017lvf,Liu:2021nyg,Hongo:2021ona,Weickgenannt:2022zxs}. Spin hydrodynamics builds upon the independent inclusion of spin degrees of freedom in a hydrodynamic description by adding an equation for the ST to those of the energy-momentum tensor and currents. In the quark-gluon plasma (QGP) produced in heavy-ion collisions spin hydrodynamics could be relevant to several measurements related to spin, see also the reviews~\cite{Becattini:2024uha,Becattini:2020ngo}. The description of spin dynamics is also crucial for magnetic materials and for spin electronics, see, for instance, the review~\cite{Maekawa2017-um}.
	
	The canonical ST for a Dirac fermion, which is obtained directly from Noether’s theorem, is
	\begin{equation}
		\label{eq:STCan}
		\wspt_{\rm C}^{\lambda,\mu\nu}
		= \frac{1}{2}\bar\psi\left\{\gamma^\lambda,\,\Sigma^{\mu\nu} \right\}\psi
		= \epsilon^{\rho\lambda\mu\nu} \frac{\bar\psi \gamma_\rho \gamma^5 \psi}{2}
		= \epsilon^{\rho\lambda\mu\nu} \frac{\h{j}_{5,\rho}}{2}
	\end{equation}
	with $\Sigma_{\mu\nu}=(\I/4)[\gamma_\mu,\,\gamma_\nu]$.
	It is completely antisymmetric and dual to the axial current $\h{j}_5$.
	However, there are other possible choices.
	Changing the ST can be seen as a different way of defining the angular momentum density in a material endowed with a finite angular momentum-boost charge $J^{\mu\nu}$.
	Different definitions of the ST can be given using Noether theorems~\cite{Daher:2022xon, Hattori:2019lfp,Cao:2022aku,singh2024nonuniqueness}, setting the theory in a curved background (minimal or nonminimal couplings to gravity)~\cite{BELINFANTE1939887,Floerchinger:2021uyo,Gallegos:2022jow,Gallegos:2021bzp}, employing effective Lagrangians~\cite{HILGEVOORD19631}, or constructed by hand~\cite{DeGroot:1980dk}. More generally, the ST can be changed by \emph{pseudogauge transformations} [defined later, see Eq.~\eqref{eq:PG transform}], which can map any of the mentioned prescriptions into any other one.
	However, there is no criterion for the choice of the ST, so its definition is ambiguous. An analogous ambiguity is
	also found in the definitions of the gravitational form factors for relativistic hadrons~\cite{Miller:2010nz,Leader:2013jra,Li:2024vgv} and in how to split the total angular momentum of a composite particle, such as the proton, into separate quark and gluon contributions, and into spin and orbital components~\cite{Liu:2011vcs,Leader:2013jra,Brambilla:2014jmp}. 
	
	In this work, we consider a plasma of quarks described by a modified Nambu-Jona-Lasinio (NJL) model~\cite{Nambu:1961tp,Ebert:1985kz} with Lagrangian
	\begin{equation}
		\label{eq:LNJL}
		\begin{split}
			\mathcal{L}_{\rm NJL} =& \bar{q}\left(\frac{\I}{2}\overleftrightarrow{\slashed{\de}}-\hat{m} \right)q 
			+ G_A \left(\bar{q}\gamma_{\mu} \gamma^5 q\right)\left(\bar{q}\gamma^{\mu}\gamma^5 q\right)+\cdots,
		\end{split}
	\end{equation}
	where the flavors of the quarks are contained in $q$, and $\hat{m}$ is the mass matrix of the quarks in flavor space,
	and additional interaction terms, e.g., $(\bar{q}\,q)^2$, are omitted. The relevance of this model for QGP physics can be justified by functional renormalization group methods, which reveal that NJL-type interactions are generated by the QCD dynamics \cite{Braun:2011pp,Braun:2017srn,Braun:2018bik,Braun:2019aow}.
	The quartic interaction through the iso-singlet axial current $\h{j}_5^\mu=\bar{q}\gamma^\mu\gamma_5 q$ can also be seen as (canonical) spin-spin interactions thanks to Eq.~(\ref{eq:STCan}).
	Therefore, this interaction will be used as an effective description of spin-spin quark interactions  in finite-temperature QCD, and will be the only interaction relevant for the present investigation.
	
	We prove that the effects of spin-spin interactions in the presence of rotation and/or magnetic field are, in the mean field approximation,  equivalent to those of the canonical ST in a theory of ideal spin hydrodynamics. This suggests that a canonical ST is dynamically generated in the QGP through the chiral separation effect (CSE) and the axial vortical effect (AVE) \cite{Kharzeev:2015znc}, identifying a mechanism that could favor a theory of spin hydrodynamics with the canonical ST. We conclude by clarifying the distinction between a fundamental definition of the ST relevant to gravitational physics and a phenomenological definition in effective theories such as hydrodynamics. We also discuss analogies between the inclusion of a ST in hydrodynamics and the Landau theory for magnetization, and conjecturing a possible way to observe the effects of spin hydrodynamics unambiguously.
	
	We use natural units where $c=\hbar=k_{\rm B}=1$.
	
	%*********************************************************************************************************
	\vskip0.3cm
	{\it{Spin tensor and spin potential-}}
	%*********************************************************************************************************
	In special relativity, the conserved currents associated with translation and Lorentz invariance are the energy-momentum tensor $\wT^{\mu\nu}$ and the total angular momentum-boost operator density
	\begin{equation}
		\label{eq:AMO}
		\h{\mathcal{J}}^{\lambda,\mu\nu} = x^\mu \wT^{\lambda\nu} - x^\nu \wT^{\lambda\mu}
		+\wspt^{\lambda,\mu\nu},
	\end{equation}
	where $\wspt$ is the \emph{spin tensor}, which is antisymmetric in the last two indices. A \emph{pseudogauge} (PG) transformation is defined as ~\cite{Hehl:1976vr}
	\begin{subequations}\label{eq:PG transform}
		\begin{align}
			\wT^{\prime\mu\nu}= &\wT^{\mu\nu}+\frac{1}{2}\nabla_\lambda\left(\wPhi^{\lambda,\mu\nu}-\wPhi^{\mu,\lambda\nu} -\wPhi^{\nu,\lambda\mu}\right),\\
			\wspt^{\prime\lambda,\mu\nu}=&\wspt^{\lambda,\mu\nu}-\wPhi^{\lambda,\mu\nu}+\nabla_\rho \widehat{Z}^{\mu\nu,\lambda\rho},
		\end{align}
	\end{subequations}
	where $\wPhi$ and $\widehat{Z}$ are arbitrary tensors obeying $\wPhi^{\lambda,\mu\nu}=-\wPhi^{\lambda,\nu\mu}$ and $\h{Z}^{\mu\nu,\lambda\rho}=-\h{Z}^{\nu\mu,\lambda\rho}=-\h{Z}^{\mu\nu,\rho\lambda}$.
	PG transformations represent an ambiguity in the definition of $\wT^{\mu\nu}$ and $\h{\mathcal{S}}^{\lambda,\mu\nu}$ because the transformed tensors obey $\nabla_\mu \wT'^{\mu\nu}=\nabla_\mu \wT^{\mu\nu}=0$ and $\nabla_\lambda \h{\mathcal{J}'}^{\lambda,\mu\nu}=\nabla_\lambda \h{\mathcal{J}}^{\lambda,\mu\nu}=0$, so the total momentum and angular momentum-boost operators are not affected by the transformation.

	Different PGs describe different energy-momentum and angular momentum densities, but the difference appears only as a quantum correction~\cite{Becattini:2020riu}. Yet, such a difference becomes extremely relevant to describe spin, which is purely a quantum observable. Indeed, in the context of heavy-ion collisions, the measurement of the $\Lambda$ spin polarization spurred the development of spin hydrodynamics, which describes the macroscopic effects of spin by including a ST and a \emph{spin potential} in the equations of relativistic hydrodynamics. In this framework, different STs lead to different spin-hydrodynamic theories and conclusions~\cite{Becattini:2011ev,Becattini:2018duy, Becattini:2020riu}. In fact, the relation of polarization with thermal vorticity, thermal shear, and spin potential is PG dependent \cite{Buzzegoli:2021wlg}.
	This observation leads to questioning which ST is physical or how to choose it \cite{Becattini:2018duy,Speranza:2020ilk,Fukushima:2020ucl,Das:2021aar,Buzzegoli:2021wlg,Daher:2022xon}. Indeed, different versions of spin hydrodynamics have been proposed based on different STs: the Belinfante \cite{Fukushima:2020ucl}, the canonical \cite{Hattori:2019lfp,Hongo:2021ona,Dey:2023hft}, the de Groot-van Leeuwen-van Weert \cite{DeGroot:1980dk,Florkowski:2018fap}, and the Hilgevoord-Wouthuysen \cite{Hilgevoord:1965,Weickgenannt:2022zxs}. In our opinion, no decisive argument favoring any PG over the others has yet been given.  We will try to fill this gap.
	
	A relativistic theory of spin hydrodynamics based on an underlying microscopic quantum field theory can be obtained with the Zubarev formalism for the non-equilibrium statistical operator $\wrho$~\cite{Hosoya:1983id,Becattini:2014yxa,Becattini:2019dxo}. The covariant form of $\wrho$ is obtained by maximizing the total entropy for fixed energy-momentum and boost-angular momentum densities~\cite{Zubarev:1966,Zubarev:1979,vanWeert1982,Zubarev:1989su,Morzov:1998}. These constraints boil down to the two equations
	\begin{align}
		n_\lambda \tr\left[\wrho\, \wT^{\lambda\nu} \right]&= n_\lambda T^{\lambda\nu},\quad
		n_\lambda \tr\left[\wrho\, \wspt^{\lambda,\mu\nu} \right]&= n_\lambda \spt^{\lambda,\mu\nu},
		\label{eq:STConstrain}
	\end{align}
	where $n$ is the normal vector to the spacelike hypersurface $\Sigma$ where local equilibrium is achieved. 
	The result is~\cite{Becattini:2018duy,Speranza:2020ilk}
	\begin{align}
		\label{eq:rhoSTC}
		\wrho=&\frac{1}{\mathcal{Z}}\exp\left[-\int_{\Sigma}\di\Sigma\;n_\lambda\wT^{\lambda\nu}\beta_\nu+\int_{\Sigma}\di\Sigma\;n_\lambda\frac{\SP_{\mu\nu}}{2}\wspt^{\lambda,\mu\nu}\right]
	\end{align}
	where $\beta$ is the four-temperature $\beta^\mu=u^\mu/T$ and $\SP$ is the \emph{spin potential}, which is an antisymmetric tensor used as Lagrange multiplier to impose the constraint associated with the average ST in Eq.~(\ref{eq:STConstrain}). In Eq.~\eqref{eq:rhoSTC} we neglect the presence of other conserved charges, but their inclusion would not change our results. Since the ST and the energy-momentum tensor operators on the lhs of Eq.~(\ref{eq:STConstrain}) are PG dependent, the statistical operator, and the quantities derived thereof, are generally PG dependent too~\cite{Becattini:2018duy}. 
	
	At global equilibrium, the spin potential must be equal to the constant thermal vorticity $\varpi_{\mu\nu}=-(1/2)(\nabla_\mu\beta_\nu-\nabla_\nu\beta_\mu)$~\cite{Becattini:2018duy,Speranza:2020ilk}. When the spin potential deviates from vorticity, the statistical operator describes a hydrodynamic theory where spin degrees of freedom are independent of standard hydrodynamic quantities~\cite{Hongo:2021ona}, and one would be able to describe phenomena like the spin polarization of (anti)particles even in a nonvortical fluid~\cite{Becattini:2020riu}.
	
	%*********************************************************************************************************
	\vskip0.3cm
	{\it{Partition function with spin potential-}}
	%*********************************************************************************************************
	We now derive the partition function of the statistical operator (\ref{eq:rhoSTC}) for a free Dirac field in the canonical PG, meaning that the operators in (\ref{eq:rhoSTC}) are the canonical ones.
	It is convenient to rewrite the (canonical) statistical operator in terms of the Belinfante energy-momentum tensor $\wT^{\mu\nu}_B$. One  obtains~\cite{Becattini:2018duy,Buzzegoli:2021wlg}
	\begin{equation}
		\widehat{\rho}=\frac{1}{\mathcal{Z}}\exp\left\{\!-\!\int_{\Sigma}\!\!\di\Sigma\, n_\mu \left[ \wT^{\mu\nu}_B\beta_\nu+\frac{1}{2}(\varpi-\SP)_{\rho\sigma}\wspt_{\rm C}^{\mu,\rho\sigma}\right]\right\}.
	\end{equation}
	The partition function of this operator can be written as a Euclidean path integral in an emergent thermal spacetime constructed from the thermodynamic fields as described in~\cite{Hayata:2015lga,Hongo:2016mqm}.  For instance, in the case of a rotating fluid, the thermal spacetime is described by rotating coordinates, as used in several studies \cite{Ambrus:2014uqa,Chen:2015hfc,Mameda:2015ria,Buzzegoli:2022dhw}.
	
	The geometry of the spacelike hypersurface $\Sigma$ embedded in the (flat) background spacetime and parametrized by a constant ``time'' coordinate $\bar{t}(x)$ and space coordinates $\bar{\bm{x}}(x)$, can be described with the Arnowitt-Deser-Misner (ADM) metric constructed with the ADM decomposition $t^\mu=N n^\mu + N^\mu$, where $t^\mu=\de_{\bar{t}}x^\mu(\bar{t},\bar{\bm{x}})$ is the local direction of time. The transformation to new thermal coordinates $\tilde{x}$, such that $t^\mu(x)=\beta^\mu(x)/B$, with $B$ a scale for the inverse temperature, defines the induced thermal metric $g_{\tilde{\mu}\tilde{\nu}}$ and the vielbein $e_{\tilde{\mu}}^{\hphantom{\tilde{\mu}}a}$ used to describe the Dirac field in a curved background, as
	\begin{equation}
		g_{\tilde{\mu}\tilde{\nu}} = \eta_{\rho\sigma} \frac{\de x^\rho}{\de \tilde{x}^{\tilde{\mu}}} \frac{\de x^\sigma}{\de \tilde{x}^{\tilde{\nu}}}
		= e_{\tilde{\mu}}^{\hphantom{\tilde{\mu}}a} e_{\tilde{\nu}}^{\hphantom{\tilde{\nu}}b} \eta_{ab}\, .
	\end{equation}
	Integrals over the constant time hypersurface are then written as
	\begin{equation}
		\label{eq:Ints}
		\int_{\Sigma}\di\Sigma\; n_\lambda= \int\di^3\tilde{\bm{x}}\,\sqrt{\tilde{g}}\tilde{N}^{-1}\, n_{\tilde{\lambda}}.
	\end{equation}
	Consequently, the partition function is obtained as the path integral of the Euclidean action of the field theory~\cite{Hayata:2015lga,Hongo:2016mqm}:
	\begin{equation}\label{eq:PI local eq}
		\mathcal{Z}=\int \mathcal{D}[\psi,\psi^\dagger] \e^{-S_E-S_\Theta},
	\end{equation}
	where $S_E$ is the Euclidean action with imaginary time $\tau$ and Euclidean emergent thermal spacetime $\tilde{g}_E$, and $S_\Theta$ is a ``spin action'':
	\begin{subequations}\label{eq: action and spin action}
		\begin{align}
			S_E&=\int_0^{B}\di\tau\di^3\tilde{\bm{x}}\sqrt{\tilde{g}_E}\left[\bar{\psi}\left(\frac{\overleftrightarrow{\slashed{D}}}{2}+m\right)\psi+V(\bar{\psi},\psi)\right],\label{eq:euclidean dirac action}\\
			S_\Theta &= \frac{1}{2}\int_0^{B}\di\tau\di^3\tilde{\bm{x}}\sqrt{\tilde{g}_E}\;n_\tilde{\mu}\Theta_{\tilde{\rho}\tilde{\sigma}}\wspt_{\rm C}^{\tilde{\mu},\tilde{\rho}\tilde{\sigma}},\label{eq: eucledian spin action}
		\end{align}
	\end{subequations}
	where we defined $\Theta=(\varpi-\SP)$ and we are using Euclidean gamma matrices.
	The external covariant derivative is $\overrightarrow{D}_{\tilde0}=\partial_{\tilde0}+(1/2)\varpi_{\tilde0}^{\hphantom{\tilde0}ab}\Sigma_{ab}$, $\overleftarrow{D}_{\tilde0}=\partial_{\tilde0}-(1/2)\varpi_{\tilde0}^{\hphantom{\tilde0}ab}\Sigma_{ab}$, and $\overrightarrow{D}_{\tilde i}=\overleftarrow{D}_{\tilde i}=\partial_{\tilde i}$ and we defined $\overleftrightarrow{\slashed{D}} = e_a^{\tilde\mu}\gamma^a\overrightarrow{D}_{\tilde\mu}-\overleftarrow{D}_{\tilde\mu}e_a^{\tilde\mu}\gamma^a$. The spin connection $\varpi$ of the thermal spacetime is related to the thermal vorticity (in the new coordinates) as~\cite{Hongo:2016mqm}
	\begin{equation}
		\label{eq:thermalgauge}
		\varpi_{\tilde{0}\tilde{i}\tilde{j}} = -\frac{1}{2B}\left(\de_{\tilde{i}}\beta_{\tilde{j}} - \de_{\tilde{j}}\beta_{\tilde{i}} \right).
	\end{equation}

	For the spin action, due to antisymmetry, we can rewrite
	\begin{equation}
		\frac{1}{2}S^{\tilde{\lambda},\tilde{\mu}\tilde{\nu}}_{\rm C} n_{\tilde\lambda}\Theta_{\tilde{\mu}\tilde{\nu}} =
		\frac{1}{6}S^{\tilde{\lambda},\tilde{\mu}\tilde{\nu}}_{\rm C}(n_{\tilde\lambda}\Theta_{\tilde{\mu}\tilde{\nu}}-n_{\tilde\mu}\Theta_{\tilde{\lambda}\tilde{\nu}}-n_{\tilde\nu}\Theta_{\tilde{\mu}\tilde{\lambda}}),
	\end{equation}
	where we have made the tensor contracting $S^{\tilde{\lambda},\tilde{\mu}\tilde{\nu}}_{\rm C}$ completely antisymmetric. Therefore one can introduce a pseudovector field $f^{\tilde\sigma}$ such that
	\begin{equation}
		\frac{1}{6}(n_{\tilde\lambda}\Theta_{\tilde{\mu}\tilde{\nu}}-n_{\tilde\mu}\Theta_{\tilde{\lambda}\tilde{\nu}}-n_{\tilde\nu}\Theta_{\tilde{\mu}\tilde{\lambda}})=\frac{1}{6}\epsilon_{\tilde\sigma\tilde\lambda\tilde\mu\tilde\nu}f^{\tilde\sigma}
	\end{equation} 
	which is solved for
	\begin{equation}
		\label{eq:rewriting of potential}
		f^{\tilde\sigma} = \frac{1}{4}\epsilon^{\tilde\mu\tilde\nu\tilde\rho\tilde\sigma}\Theta_{\tilde{\mu}\tilde{\nu}}n_{\tilde\rho}
		= \frac{1}{4}\epsilon^{\tilde\mu\tilde\nu\tilde\rho\tilde\sigma}(\varpi_{\tilde{\mu}\tilde{\nu}}-\SP_{\tilde{\mu}\tilde{\nu}})n_{\tilde\rho},
	\end{equation}
	where $\epsilon$ is the Levi-Civita tensor.
	Furthermore, using the above equation and the duality \eqref{eq:STCan}, the Eq. \eqref{eq: eucledian spin action} reads
	\begin{equation}\label{eq:spin action with j5}S_\Theta=-\int_0^{B}\di\tau\di^3\tilde{\bm{x}}\sqrt{\tilde{g}_E}\; j_5^{\tilde{\mu}} f_\tilde{\mu}.
	\end{equation}
	This observation will be useful later on.
	
	A geometrical interpretation of the spin potential is obtained by rewriting the spin action as follows
	\begin{equation}
		\label{eq:PISP}
		\begin{split}
			S_{\Theta} =& \int_{0}^{B}\!\!\!\!\di\tau\di^3\tilde{\bm{x}}\,\sqrt{\tilde{g}_E}\,
			S_{\rm C}^{\tilde{\lambda},\tilde{\mu}\tilde{\nu}} \frac{1}{2}n_{\tilde\lambda}\Theta_{\tilde{\mu}\tilde{\nu}}\\
			=& \frac{1}{4}\int_{0}^{B}\!\!\!\! \di\tau\di^3\tilde{\bm{x}}\,\sqrt{\tilde{g}_E}\,e_a^{\tilde{\lambda}}
			\bar\psi\left\{ \gamma^a,\,\Sigma_{bc}\right\}\psi\, \omega_{\tilde\lambda}^{\hphantom{\tilde\lambda}bc}\\
			= & \frac{1}{2}\int_{0}^{B}\!\!\!\! \di\tau\di^3\tilde{\bm{x}}\,\sqrt{\tilde{g}_E}\,\bar{\psi}\left\{e_a^{\tilde{\mu}}\gamma^a,\, \SPd_{\tilde{\mu}} \right\}\psi\,,  
		\end{split}
	\end{equation}
	where we defined
	\begin{equation}
		\omega_{\tilde\lambda}^{\hphantom{\tilde\lambda}bc}=\frac{1}{2}e^b_{\tilde{\mu}}e^c_{\tilde{\nu}}n_{\tilde\lambda}\Theta^{\tilde{\mu}\tilde{\nu}},\quad
		\SPd_{\tilde{\mu}} =  \tfrac{1}{2}\omega_{\tilde{\mu}ab}\Sigma^{ab}.
	\end{equation}
	Note that in the thermal coordinates the only nonvanishing components of $\omega_{\tilde\lambda}^{\hphantom{\tilde\lambda}bc}$ have $\tilde\lambda=0$.
	Comparing with Eq.~\eqref{eq:euclidean dirac action}, the integrand of $S_\Theta$ can be reabsorbed in the Lagrangian by changing the covariant derivative $D_{\tilde{\mu}} \to D_{\tilde{\mu}} +\SPd_{\tilde{\mu}}$, meaning that the total spin connection becomes $\omega'=\varpi+\Theta=2\varpi-\SP$, acquiring a part which is independent of the metric.  This is in agreement with the proper definition of the canonical ST in general relativity being obtained in space times with torsion~\cite{Speranza:2020ilk,Hongo:2021ona,Gallegos:2021bzp} (Einstein-Cartan theory~\cite{Hehl:1976kj,Hehl:1984ev,Hehl:2007bn,Blagojevic2001}). 
	For $\Theta=0$ the connection equals the thermal vorticity, so the global equilibrium with the canonical PG coincides with the Belinfante PG, as it should be.
	
	%*********************************************************************************************************
	\vskip0.3cm
	{\it{Spin potential generation-}}
	%*********************************************************************************************************
	We now show that the same partition function as in Eq. \eqref{eq:PI local eq} is obtained from the modified NJL model in Eq.~(\ref{eq:LNJL}) at finite temperature and rotation.
	The NJL partition function in the rotating plasma is written as the Euclidean path integral in the thermal metric background, that is
	\begin{equation}
		\label{eq:ZNJL}
		\mathcal{Z}_{\rm NJL} = \int\mathcal{D}[q,q^\dagger]\,
		\e^{-S_{\rm NJL}[q,\,\bar{q};\,e^{\tilde\mu}_a]},
	\end{equation}
	where the action is written as in Eq.~(\ref{eq:euclidean dirac action}) with the NJL Lagrangian~(\ref{eq:LNJL}). 
	
	We follow the standard approach~\cite{Ebert:1985kz} and we make a Hubbard-Stratonovich transformation to introduce a pseudovector field $f_\mu$, corresponding to the quark-antiquark singlet axial current:
	\begin{equation}
		\mathcal{Z}_{\rm NJL} = \int\mathcal{D}[q,q^\dagger]\mathcal{D}f\,
		\e^{ -S_f[q,\,\bar{q},\,f_{\tilde{\mu}};\,e^{\tilde\mu}_a]}
	\end{equation}
	with
	\begin{equation}
		\label{eq:Sf}
		\begin{split}
			S_f \!=& \!\!\int_0^{B}\!\!\!\!\!\di\tau\!\!\int\!\!\di^3\tilde{\bm{x}}\sqrt{\tilde{g}_E}\left[
			\bar{q}(\frac{\I}{2}\overleftrightarrow{\slashed{D}}-\hat{m})q
			+\!\frac{1}{4G_A}f_{\tilde\mu}f^{\tilde\mu}\!+\!\mathcal{L}_{f}\right]
		\end{split}
	\end{equation}
	where $\mathcal{L}_{f}$ reads
	\begin{equation}\label{eq:int NJL}
		\mathcal{L}_f = -{j_5}_{\tilde{\sigma}} f^{\tilde\sigma}=\frac{1}{3}e_a^{\tilde{\lambda}}S_{C}^{a,bc}e_b^{\tilde{\mu}} e^{\tilde{\nu}}_{c}\, \epsilon_{\tilde{\sigma}\tilde{\lambda}\tilde{\mu}\tilde{\nu}} f^{\tilde\sigma},
	\end{equation}
	where we have used the duality between the axial current and the canonical ST.
	
	Comparing Eq. \eqref{eq:int NJL} with Eq. \eqref{eq:spin action with j5}, we realize that the field $f$ generates the same action as the canonical spin potential in spin hydrodynamics.
	The equivalence between the models is fully realized in the mean-field approximation, where the path integral over $f$ yields the substitution of the dynamical field $f$ with its mean value
	\begin{equation}
		\int\!\!\mathcal{D}[q,\bar{q}]\!\! \int\!\!\mathcal{D}f\,
		\e^{-S_f[q,\bar{q},f_{\tilde\mu};\,e^{\tilde\mu}_a]}\mapsto
		\int\!\!\mathcal{D}[q,\bar{q}]\e^{-S_f[q,\bar{q},\langle f_{\tilde\mu}\rangle;\,e^{\tilde\mu}_a]}.
	\end{equation}
	In this case, according to the correspondence \eqref{eq:rewriting of potential}, the partition function $\mathcal{Z}_{\rm NJL}$ coincides with Eq.~\eqref{eq:PI local eq} up to a constant term.
	Hence, the mean field $\langle f\rangle$ describes the difference between thermal vorticity and spin potential.
	This proves that the interacting theory given by Eq.~\eqref{eq:LNJL} in the mean field approximation is equivalent to a free theory at local equilibrium with the canonical ST.
	
	Physically, a mean $f_\mu$ is expected if the state explicitly breaks parity and Lorentz symmetry, such that the singlet axial current $\langle \bar{q}\gamma_\mu\gamma^5 q\rangle$ is nonvanishing. This is the case if magnetic field at finite charge density and/or rotation are present: an axial current is induced by magnetic field and rotation by the CSE and the AVE \cite{Kharzeev:2015znc}. We expect that both the CSE and the AVE will lead to a spin potential generation.  In the low energy limit, the field $f_\mu$ emerging in this way from the NJL model represents the isospin-singlet axial mesons. Because of charge conjugation properties, $f_\mu$ has the same quantum numbers as the $f_1$ or $h_1$ mesons, depending on whether the field is induced by rotation or magnetic fields. 
	
	We stress that the CVE and the CSE do not require a chiral imbalance and local $CP$ violations. The CSE has been observed in first principle lattice QCD calculations~\cite{Brandt:2023wgf}, where it was shown that the CSE conductivity is non vanishing in the hot chiral symmetric phase of QCD. 
	A proper study of spin-spin interactions in hot QCD can be done through the renormalization group flow, repeating the analysis of Refs.\cite{Braun:2011pp,Braun:2017srn,Braun:2018bik,Braun:2019aow} in the presence of rotation and/or magnetic field. From this observation, we expect that the ST emerging from the QGP remains the canonical one.
	However, more detailed studies are in order.
	
	We point out that the derivation presented here is valid even if the magnetic field and the rotation are vanishing. However, one would need to provide a different mechanism to justify a nonvanishing mean axial vector field $f_\mu$.
	%*********************************************************************************************************
	\vskip0.3cm
	{\it{Discussion-}}
	%*********************************************************************************************************
	The dynamical generation of ST in the NJL model described above is reminiscent of the emergence of magnetization in ferromagnetism. Considering an Ising model with infinite range interactions given by a Hamiltonian of $N$ spins with interactions $\h{H}=-(J/2N)\sum_{i,j}S_i\,S_j$ \cite{Goldenfeld2018}, the Hubbard-Stratonovich identity yields the statistical operator 
	\begin{equation}
		\wrho=\sqrt{\frac{N\beta J}{2\pi}}\int_{-\infty}^\infty \di\mu \exp\left[-\frac{N\beta J}{2}\mu^2+\beta J\mu\h{M}\right],
	\end{equation}
	where the magnetization $\h{M}=\sum_i S_i$, coupled to the polarization $\mu$, is generated just like the ST coupled to the spin potential (\ref{eq:rewriting of potential}) in the NJL model.
	
	Furthermore, spin hydrodynamics can be interpreted as an analog to the phenomenological Landau theory for magnetization. Indeed, the partition function with ST can be seen as the Landau function $L$. Given a coarse-graining scale $\Lambda$ and the observation of an inhomogeneous magnetization $M_\Lambda(\bm{x})$, the Landau function is obtained from the statistical operator $\exp(-\beta \h{H})$ by fixing the magnetization, i.e., by tracing out the configurations of spins $\{S_i\}$ resulting in a different magnetization as follows~\cite{Goldenfeld2018}:
	\begin{equation}
		\begin{split}
			\e^{-\beta L\left\{ M_\Lambda(\bm{x})\right\}} = & \tr\left\{ \e^{-\beta \h{H}\{S_i\}}
			\delta\left[\sum_{i \in \bm{x}} S_i - M_\Lambda(\bm{x}) \right] \right\}.
		\end{split}
	\end{equation}
	Similarly, in the relativistic spin hydrodynamics we obtained the statistical operator by imposing the constraint in Eq.~(\ref{eq:STConstrain}). This suggests that we can use spin hydrodynamics as a phenomenological theory and choose the ST that best describes the spin properties of the system at the desired scale. In fact, the magnetization itself is subject to PG ambiguity. In a relativistic context, the conserved electric current can be redefined as
	\begin{equation}
		\h{j}^{\prime\mu} = \h{j}^\mu + \nabla_{\lambda}\h{\mathcal{M}}^{\lambda\mu},
		\quad \h{\mathcal{M}}^{\lambda\mu} = -\h{\mathcal{M}}^{\mu\lambda}, 
	\end{equation}
	where $\nabla_{\lambda}\h{\mathcal{M}}^{\lambda\mu}$ is interpreted as a bound current and the arbitrary magnetization $\h{\mathcal{M}}$ is chosen to best describe the material. This analogy indicates that for a macroscopic spin hydrodynamics theory, one can give up a fundamental interpretation of the ST, and instead use the ST and the spin potential as an effective description of emerging spin properties of the system. In this connotation, the PG dependence is not a problem to be solved. However, in the absence of nonvanishing boundary terms, these currents are impossible to detect.
	In this work, we showed that the presence of the ST in the statistical operator can be derived from an underlying microscopic theory. In the case study, axial current interactions in the NJL model generate a canonical ST, but different models could result in different PGs.
	
	The fact that observable quantities, such as the spin polarization measured in heavy-ion collisions~\cite{STAR:2017ckg}, depend on the choice of the ST~\cite{Buzzegoli:2021wlg} is a consequence of the PG dependence of the statistical operator (\ref{eq:rhoSTC}), which is ultimately caused by coarse-graining and our ignorance about the system, see Ref.~\cite{Becattini:2020riu}.
	The PG dependence is also present in Wigner functions, Kubo formulas, and quantum kinetic methods because they are all equivalent.
	Moreover, the PG ambiguity in momentum-dependent quantities, such as the average spin polarization, implies that PG dependence cannot be reabsorbed into a redefinition of the thermodynamic fields~\cite{Buzzegoli:2021wlg}.
	
	One may wonder if it is currently possible to detect a spin potential and the onset of spin hydrodynamics in heavy-ion collisions. Indeed, an attempt to detect a spin potential from the available polarization data of heavy-ion collisions seems difficult. The reason is that the Belinfante PG, where the ST is vanishing, gives a good description of the polarization data, especially since the theoretical predictions depend rather strongly on the details of initial conditions, transport coefficients, and equation of state  \cite{Ryu:2021lnx,Alzhrani:2022dpi,Wu:2022mkr,Palermo:2024tza}. 
	
	A clear-cut evidence of a spin hydrodynamics regime would be the observation of spin polarization in the absence of vorticity, as explained in Ref. \cite{Becattini:2020riu}. An ideal realization of this example would be implemented in heavy-ion collisions by creating an initial state where most of the angular momentum is carried by spin degrees of freedom, e.g., by polarizing heavy ions before performing very central collisions. In this case, a possible signal of polarization could be only explained by invoking a hydrodynamic transport of spin in the QGP through spin hydrodynamics. However, the realization of polarized beams of heavy ions seems difficult, and the angular momentum present in the initial state would be small in any case. Nonetheless, the study of polarization in very central collisions, where the effect of vorticity is minimized, could still give indications on the necessity of including spin in relativistic hydrodynamics, regardless of the polarization state of the colliding beams.
	
	%*********************************************************************************************************
	\vskip0.3cm
	{\it{Conclusions-}}
	%*********************************************************************************************************
	Pseudogauge symmetry curses relativistic spin hydrodynamics. In this Letter, we have proposed a new interpretation where a microscopic mechanism based on quantum field theory selects the form of the spin tensor. 
	
	We have analyzed the spin-spin interactions in the NJL model, explicitly proving that the partition function of this theory, in the mean field approximation, is equivalent to the one of a free theory at local equilibrium with a canonical ST. In this framework, the mean axial field induces a difference between spin potential and vorticity. An equivalence of this kind may exist also in QCD, and more studies in this sense are desirable. Going beyond the tree level and the mean-field approximation, as well as performing a similar study as in Ref. \cite{Braun:2019aow} will be objects of future studies.
	
	In heavy-ion collisions, the mean axial field required for the spin potential generation is induced by the explicit parity and Lorentz symmetry breaking due to rotation or magnetic field, namely the AVE and the CSE.  Spontaneous symmetry breaking of the Lorentz and $CPT$ symmetry, introduced in extensions of the standard model \cite{Kostelecky:1988zi,Kostelecky:1991ak,Colladay:1996iz,Colladay:1998fq,Coleman:1998ti}, might also induce a mean field as described here.
	
	In conclusion, our analysis favors the use of the canonical ST in relativistic spin-hydrodynamics theory for heavy-ion collisions.
	%%%%%%%%%%%%%%%%%%%%%%%%%%%%%%%%
	\bigskip
	\acknowledgments
	{\it{Acknowledgments-}}
	M.B. is grateful to K. Tuchin for many fruitful discussions.
	A.P. acknowledges useful discussions with S. Bhadury, D. Kharzeev, F. Murgana, and M. Stephanov. We are also grateful to A. Abanov, F. Becattini, E. Grossi, and D. Teaney.
	Our work was supported by the U.S. Department of Energy under Grants No.\ DE-SC0023692 (M.B.) and DE-FG88ER40388 (A.P.) and partially by (M.B.) the European Union - NextGenerationEU through grant No. 760079/23.05.2023, funded by the Romanian Ministry of Research, Innovation and Digitization through Romania's National Recovery and Resilience Plan, call No. PNRR-III-C9-2022-I8.

\end{document}